# Adaptive Layered Approach using Machine Learning Techniques with Gain Ratio for Intrusion Detection Systems

Heba Ezzat Ibrahim
Arab Academy for Science, Technology and Maritime Transport
Cairo, Egypt

Sherif M. Badr
Arab Academy for Science, Technology and Maritime Transport
Cairo, Egypt

Mohamed A. Shaheen
Arab Academy for Science, Technology and Maritime Transport
Alexandria, Egypt

## ABSTRACT
Intrusion Detection System (IDS) has increasingly become a crucial issue for computer and network systems. Optimizing performance of IDS becomes an important open problem which receives more and more attention from the research community. In this work, A multi-layer intrusion detection model is designed and developed to achieve high efficiency and improve the detection and classification rate accuracy .we effectively apply Machine learning techniques (C5 decision tree, Multilayer Perceptron neural network and Naïve Bayes) using gain ratio for selecting the best features for each layer as to use smaller storage space and get higher Intrusion detection performance. Our experimental results showed that the proposed multi-layer model using C5 decision tree achieves higher classification rate accuracy, using feature selection by Gain Ratio, and less false alarm rate than MLP and naïve Bayes. Using Gain Ratio enhances the accuracy of U2R and R2L for the three machine learning techniques (C5, MLP and Naïve Bayes) significantly. MLP has high classification rate when using the whole 41 features in Dos and Probe layers.

## 1. INTRODUCTION
The rapid development and expansion of World Wide Web and local network systems have changed the computing world in the last decade. The costs of temporary or permanent damages caused by unauthorized access of the intruders to computer systems have urged different organizations to increasingly implement various systems to monitor data flow in their networks [1]. These systems are generally referred to as Intrusion Detection Systems (IDSs).

Intrusion Detection Systems have gained acceptance as a necessary addition to every organization’s security infrastructure [2]. Intrusion detection systems are classified as host based or network based IDS. A host based IDS defines the patterns that are detected in the event log records and monitor resources such as system logs, file systems and disk resources while a network based intrusion detection system monitors the data passing through the network when the system is used to analyze network packets. Different detection techniques can be employed to search for attack patterns in the monitored data but Host-based and network-based systems are both required because they provide significantly different benefits [3].
There are two main approaches to the design of IDSs. In a misuse detection based IDS (signature-based detection), intrusions are detected by looking for activities that correspond to known signatures of intrusions or vulnerabilities. On the other hand, an anomaly detection based IDS detects intrusions by searching for abnormal network traffic [4].

This work aims to design and develop security architecture (intrusion detection and prevention system) for computer networks. We build the model to improve the classification rate for known and unknown attacks with minimum number of false alarm rate. We train and test our proposed model on the normal and the known attacks. Then we test our system for unknown attacks by exposing the system to new attacks' types that are never seen before by the training module.

Our Model should be placed at the network server to monitor all passing data packets and determine suspicious connections. Therefore, it can alarm the system administrator with the malicious attack type. Moreover, the proposed system allows new attack types to be defined, i.e. the proposed system should have an adaptive capability.

## 2. RELATED WORK
Intrusion Detection Systems (IDS) are software or hardware tools that automatically examine, check and observe events that take place in a computer or a network, looking for indication of intrusion [5].
Intrusion detection (ID) is a major research problem in network security, where the concept of ID was proposed by Anderson in 1980 [6]. ID is based on the assumption that the behavior of intruders is different from a legal user [7]. The goal of intrusion detection systems (IDS) is to identify unusual access or attacks and raises an alarm whenever a suspicious activity is detected to secure internal networks [8]

Several machine-learning techniques including neural networks, fuzzy logic [9], support vector machines (SVM) [6, 10] have been studied for the design of IDS. In particular, these techniques are developed as classifiers, which are used to classify whether the incoming network traffics are normal or an attack.

There are researches that implement an IDS using Multilayer perceptron (MLP) which have the capability of detecting normal and attacks connection as in [11], [12].
Reference [4] used MLP not only for detecting normal and attacks connection but also identify attack type.
Decision Tree (C4.5 Algorithm) was explored as intrusion detection models in [13] and [14].

Debar et al. [15] and Zhang et al. [16] discuss the use of artificial neural networks for network intrusion detection.
Though the neural networks can work effectively with noisy data, they require large amount of data for training and it is often hard to select the best possible architecture for a neural network.

Authors in [17] proposed a simple practical layered approach to intrusion be computation. They discussed that such a system would decrease computational intensive and be more accurate. They proposed a three layer system to ensure complete security viz. availability, integrity and confidentiality, each layer corresponding to one aspect of security. the first layer or the connection establishment layer corresponds to the packet level features such as source and destination IP address, number of connections to the host, source and destination port number, user ID etc. and is optimized to detect attacks exploiting the availability aspect such as DoS attacks, probes, etc. the second layer which is the privacy layer ensures data confidentiality and refers to features such as files accessed, data retrieved etc. the third layer or access control layer ensures integrity of data and is more concerned with the file modifications, user privileges etc.

Reference [18] discussed layered approach and compared the proposed Layered Approach with the decision tress, naive Bayes classification methods. Their system is based upon serial layering of multiple hybrid detectors.

We compare the layered approach with the work in [19] where it is the most closely related work to our work. The authors in [19] addressed these two issues of Accuracy and Efficiency using Conditional Random Fields and Layered Approach. They first select four layers corresponding to the four attack groups (Probe, DoS, R2L, and U2R) then train a separate model with CRFs for each layer using the feature selection. Plug in the trained models sequentially such that only the connections labeled as normal are passed to the next layer. if the instance is labeled as attack, block it and identify it as an attack represented by the layer name , Else pass the sequence to the next layer. Our Experimental results prove that The key difference between [19] and our work is that First, the authors in [19] allow the normal instance to pass through the four layers which may increase the false alarm rate and decrease the detection rate at any layer. While we detect Normal and attacks instances in the first stage then Attacks are sent for further classification to the next stage using layered approach without Normal records. Second our work has higher detection and classification Rate for New Attacks which never been seen before. Third, our work is adaptive as if errors occurred and attacks classified incorrectly at any layers and propagated to the next layer, It will be classified as unknown attacks but in [19], if errors propagated, it may detect attacks as normal if the attack had not been detected at any layer which increase the False Negative and expose the system to dangerous attack. Finally our system can identify each attack type.

# 3. MACHINE LEARNING ALGORITHMS APPLIED TO INTRUSION DETECTION

Three distinct machine learning algorithms were tested on the NSL-KDD dataset. These algorithms are C5.0 decision trees, Multi-Layer Perceptron neural networks, and Naïve Bayes.

## 3.1 C5.0 Decision Trees

Decision trees have also been used for intrusion detection [19]. The decision tree is a simple if then else rules but it is a very powerful classifier and proved to have a high detection rate. Each decision tree represents a rule which categorizes data according to these attributes. A decision tree consists of nodes, leaves, and edges [21].
See5.0 (C5.0) is one of the most popular inductive learning tools originally proposed by J.R.Quinlan as C4.5 algorithm (Quinlan, 1993) [21]. C5.0 can deal with missing attributes by giving the missing attribute the value that is most common for other instances at the same node. Or, the algorithm could make probabilistic calculations based on other instances to assign the value [22]. Single C5 acquires pruned decision tree with pruning severity 75% and winnowing attributes.

## 3.2 Multi-Layer Perceptron (MLP) Neural Networks

The neural network gains the experience initially by training the system to correctly identify pre-selected examples of the problem [11].
The most popular static network is the MLP .MLP are feed-forward neural networks trained with the standard back propagation algorithm. They are supervised networks so they require a desired response to be trained. They are widely used for pattern classification. With one or two hidden layers, they can approximate virtually any input–output map.

## 3.3 Naïve Bayes

Naive Bayes classifiers have also been used for intrusion detection [20]. However, they make strict independence assumption between the features in an observation resulting in lower attack detection accuracy when the features are correlated, which is often the case for intrusion detection.

# 4. THE PROPOSED SYSTEM

The proposed system is layered-model approach. It is divided into 2 stages. First stage detects normal and attack records. Second stage classifies the attacks detected by stage 1. Stage 2 consists of four layers. Each layer was examined with different machine learning techniques mentioned in Section 3. Our proposed system has the ability to reduce computation and time required to detect intrusive events. It also improved detection and classification rate of normal and attack records. This approach has the advantage to flag for suspicious record even if attack type of this record wasn't identified correctly.

## 4.1 Layered Approach For Intrusion Detection

The Layer-based Intrusion Detection System (LIDS) draws its motivation from what we call as the Airport Security model, where a number of security checks are performed one after the other in a sequence. Similar to this model, the LIDS represents a sequential Layered Approach and is based on ensuring availability, confidentiality, and integrity of data and (or) services over a network [17].

## 4.2 The Proposed Layered-Model Intrusion Detection System

Our system is a modular network-based intrusion detection system that analyzes Tcpdump data using data mining techniques to classify the network records to not only normal and attack but also identify attack type.

The main characteristics of our system:

- First, our system has the capability of classifying network intruders into two stages. The first stage classifies the network records to either normal or attack. The second stage consists of four sequential Layers which can identify four categories/classes and their attack type. The data is input in the first stage which identifies if this record is a normal record or attack. If the record is identified as an attack then the module would raise a flag to the administrator that the coming record is an attack then the module inputs this record to the second stage which consists of four sequential Layers, one for each class type (DOS, Probe, U2R, R2L). Each Layer is responsible for identifying the attack type of coming record according to its class type. Else the attack passes through the next layer. If attack record couldn't be classified in the four layers, it will be labeled as unknown attacks.

  The idea is that if ever the attack type or category of the second stage is misclassified then at least the admin was identified that this record is suspicious after the first stage network. Finally the admin would be alerted of the suspected attack type to guide him for the suitable attack response [23].

- Second, it takes less training time and even decrease in each layer where we use the whole dataset for training stage 1 then in stage 2 we use only the attacks for training excluding the normal records. Then each layer act as a filters that classifies the attacks of each layer category which eliminate the need of further processing at subsequent layers but we took in consideration the propagation of errors as to simulate the real system and results be more accurate and real .

- Third, we used a layered model to reduce the computation and the overall time required to detect anomalous events and attack type. Every layer is trained separately to detect each attack category and then deployed sequentially. Our model consists of four sequential layers that correspond to the four attack categories (DoS layer, Probe layer, U2R layer, and R2L layer). We implement our system with gain ratio feature selection technique for selecting the best features for each layer based on the attacks' type that the layer is trained to detect rather than using all the 41 features .In order to make the layers independent, some features may be present in more than one layer. So we can use smaller storage space and get higher Intrusion detection performance during both the training and the testing of the system as it improves the speed of the operations of the system.
  In many situations, there is a trade-off between efficiency and accuracy of the system and there can be various avenues to improve system performance [19]. We implement the Layered Approach to improve overall system performance as our layered intrusion detection model using C5.0 decision tree achieves high efficiency and improves the detection and classification rate accuracy with low false alarm rate.

- Fourth, it is Adaptive as the training module can be retrained at any point of time which makes its implementation adaptive to any new environment and/or any new attacks in the network where Attacks that are misclassified by the IDS as normal instances or given wrong attack class/type will be relabeled by the network administrator.

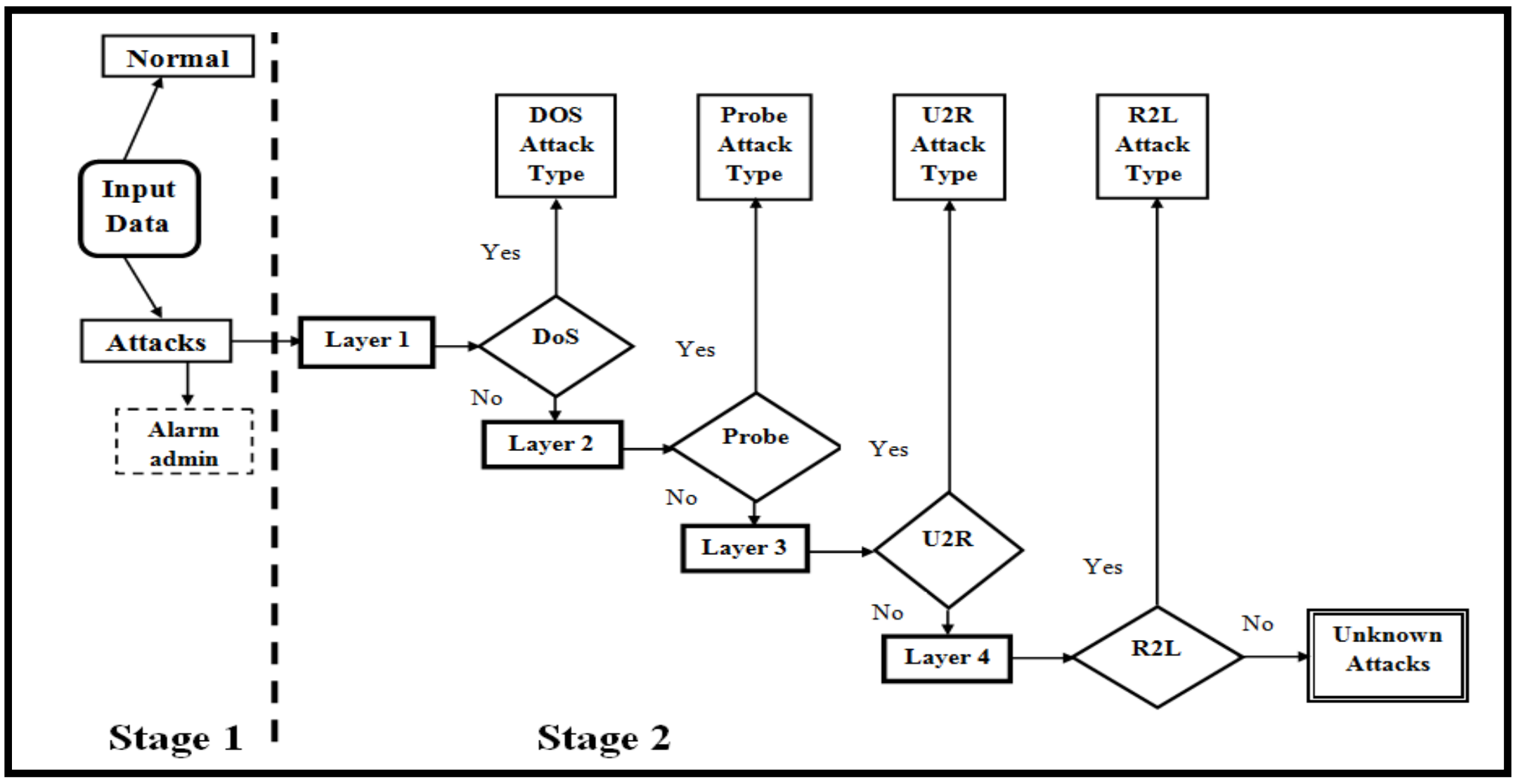

**Fig 1: The Proposed Layered-Model Approach System**

## 4.3 Gain Ratio Feature Selection

Subsequent to preprocessing of data, the features of the data set are identified as either being significant to the intrusion detection process, or redundant. This process is known as feature selection. Redundant features are generally found to be closely correlated with one or more other features. As a result, omitting them from the intrusion detection process does not degrade classification accuracy. In fact, the accuracy may improve due to the resulting data reduction, and removal of noise and measurement errors associated with the omitted features. Therefore, choosing a good subset of features proves to be significant in improving the performance of the system [24].

*Information Gain:* In this method, the features are filtered to create the most prominent feature subset before the start of the learning process.

*Gain ratio*: a modification of the information gain that solves the issue of bias towards features with a larger set of values, exhibited by information gain. Gain ratio should be Large when data is evenly spread and small when all data belong to one branch attribute.

Gain ratio takes number and size of branches into account when choosing an attribute as It corrects the information gain by taking the *intrinsic information* of a split into account (i.e. how much information do we need to tell which branch an instance belongs to) where Intrinsic information is the entropy of distribution of instances into branches.

For a given feature *x* and a feature value of *y*, it is calculated as follows:

$$Gain\ ratio(y, x) = \frac{gain(y,x)}{intrinsic\ info(x)}$$

Where,

$$Intrinsic\ info(x) = -\sum \frac{|S_i|}{|S|} * Log_2 \frac{|S_i|}{|S|}$$

$|S|$ is the number of possible values a feature $x$ can take, and $|S_i|$ is the number of actual values of feature $x$.

We select the higher gain ratio of features for each layer depending on the attacks' type that the layer is trained to detect. We select 19 features for Dos Layer, 13 features for Probe Layer, 8 Features for U2R Layer and 7 Features for R2L layer.

From our experiments done with feature selection, we have observed that Gain ratio feature selection contributed to improve overall accuracy, and improved the classification rate of instances with low frequency in the training data.

# 5. EXPERIMENTS AND RESULTS

## 5.1 Data Description

The data in the experiment is acquired from the NSLKDD dataset which consists of selected records of the complete KDD data set and does not suffer from mentioned shortcomings by removing all the repeated records in the entire KDD train and test set, and kept only one copy of each record [25]. Although, the proposed data set still suffers from some of the problems and may not be a perfect representative of existing real networks, because of the lack of public data sets for network-based IDSs, but still it can be applied as an effective benchmark data set to help researchers compare different intrusion detection methods. The NSL-KDD dataset is available at [26].

## 5.2 Performance Measure

For our results, we give the Precision, Recall, and F-Value and Accuracy. Achieving very high accuracy is very easy by carefully selecting the sample size but if we use accuracy as a measure for testing the performance of the system, the system can be biased and can attain very high accuracy. However, Precision, Recall, and F-Value are not dependent on the size of the training and the test samples.
They are defined as follows:

$$Precision = \frac{TP}{TP+FP}$$

$$Recall = \frac{TP}{TP+FN}$$

$$F\text{-}Value = \frac{(1+\beta^2)*Precision*Recall}{\beta^2*(Precision+Recall)}$$

$$Accuracy = \frac{TP+TN}{TP+TN+FP+FN}$$

Where TP, FP, FN and TN are the number of True Positives, False Positives, False Negatives and True Negative, respectively, and $\beta$ corresponds to the relative importance of precision versus recall and is usually set to 1.
We divide the training data into different groups; DoS, Probe, U2R, and R2L. Similarly, we divide the test data.

### *5.2.1 First Stage Results*

Stage 1 duty is to classify whether coming record is normal or attack. It is observed that C5 has a significant detection rate for known and unknown attacks compared to MLP and NB. The results of Stage 1 are shown in table 1 & 2.

**Table 1. Detection Rate & False Alarm Rate for Stage 1**

| Classifier | Detection Rate | False Alarm Rate |
|---|---|---|
| **C5.0** | 100 | 0 |
| **MLP** | 93.38 | 9.5 |
| **Naïve Bayes** | 98.58 | 16.78 |

**Table 2. Performance Measure for Stage 1**

| | Precision (%) | Recall (%) | F-Value (%) | Accuracy (%) |
|---|---|---|---|---|
| **C5.0** | 100 | 100 | 100 | 100 |
| **MLP** | 90.41 | 92.86 | 91.62 | 91.89 |
| **Naïve Bayes** | 83.22 | 98.21 | 90.18 | 91.1 |

### *5.2.2 Second Stage Results*

Records classified as attacks by the first Stage are introduced to second Stage which is responsible for classifying coming attack to one of the four classes (DOS, Probe, U2R and R2L) and identifying its attack type. Stage 2 consists of four sequential layers; a layer for each class which identify the class of each coming attack.
We perform two sets of experiments. From the first experiment, the systems are trained using all the 41 features. The second experiment where we perform feature selection by using Gain Ratio as to select the best features for each layer instead of using all the 41 features. We perform the same experiment with C5 decision trees, MLP and naive Bayes and compare the results.

#### *5.2.2.1 DoS Layer*

Results of Denial of service Layer with 41 Features showed that C5 decision tree has significant result. Also MLP showed promising results than naive Bayes as shown in table 3.

**Table 3. Performance Measure for Dos Layer with 41 Features**

| | Precision (%) | Recall (%) | F-Value (%) | Accuracy (%) |
|---|---|---|---|---|
| **C5.0** | 100 | 100 | 100 | 100 |
| **MLP** | 99.96 | 99.89 | 99.93 | 99.89 |
| **Naïve Bayes** | 81.3 | 76.4 | 78.8 | 74.03 |

Denial of service Layer using Gain ratio showed that C5 decision tree are more efficient than MLP and naive Bayes as shown in table 4.

**Table 4. Performance Measure for Dos Layer with Gain Ratio**

| | Precision (%) | Recall (%) | F-Value (%) | Accuracy (%) |
|---|---|---|---|---|
| **C5.0** | 100 | 100 | 100 | 100 |
| **MLP** | 97.29 | 87.34 | 92.04 | 88.97 |
| **Naïve Bayes** | 83.5 | 88.5 | 85.9 | 81.74 |

### *5.2.2.2 Probe Layer*

Results of Probe Layer with 41 features showed that C5 & MLP are most efficient for detecting this type of attacks as shown in table 5. While Probe Layer using Gain ratio showed that C5 decision tree are more efficient than MLP and naive Bayes as shown in table 6.

**Table 5. Performance Measure for Probe Layer with 41 Features**

| | Precision (%) | Recall (%) | F-Value (%) | Accuracy (%) |
|---|---|---|---|---|
| **C5.0** | 100 | 100 | 100 | 100 |
| **MLP** | 99.96 | 100 | 99.98 | 99.96 |
| **Naïve Bayes** | 81.3 | 99.6 | 89.5 | 88.95 |

**Table 6. Performance Measure for Probe Layer with Gain Ratio**

| | Precision (%) | Recall (%) | F-Value (%) | Accuracy (%) |
|---|---|---|---|---|
| **C5.0** | 100 | 100 | 100 | 100 |
| **MLP** | 74.5 | 83.49 | 78.21 | 70.92 |
| **Naïve Bayes** | 84.1 | 99.6 | 91.2 | 88.86 |

### *5.2.2.3 U2R Layer*

The U2R attacks are very difficult to detect and most of the present intrusion detection systems fail to detect such attacks with acceptable reliability. Our proposed system can be used to reliably detect such attacks. U2R Layer with 41 features showed Naïve bayes has significant higher accuracy compared to C5 and MLP as shown in table 7. While using Gain Ratio, it showed that C5 is the best classifier as shown in table 8.

**Table 7. Performance Measure for U2R Layer with 41 Features**

| | Precision (%) | Recall (%) | F-Value (%) | Accuracy (%) |
|---|---|---|---|---|
| **C5.0** | 100 | 75.41 | 85.98 | 95.18 |
| **MLP** | 82.76 | 39.34 | 53.33 | 86.50 |
| **Naïve Bayes** | 80 | 96.6 | 87.5 | 99.21 |

**Table 8. Performance Measure for U2R Layer with Gain Ratio**

| | Precision (%) | Recall (%) | F-Value (%) | Accuracy (%) |
|---|---|---|---|---|
| **C5.0** | 100 | 100 | 100 | 100 |
| **MLP** | 80 | 32.65 | 44.84 | 96.17 |
| **Naïve Bayes** | 100 | 100 | 100 | 100 |

### *5.2.2.4 R2L Layer*

Results of R2L Layer with 41 features showed that Naïve Bayes has higher accuracy rate than C5 and MLP as shown in table 9. While R2L Layer using Gain ratio showed that C5 decision tree and Naïve Bayes has significant result compared to MLP as shown in table 10.

**Table 9. Performance Measure for R2L Layer with 41 Features**

| | Precision (%) | Recall (%) | F-Value (%) | Accuracy (%) |
|---|---|---|---|---|
| **C5.0** | 95.69 | 97.6 | 96.64 | 93.58 |
| **MLP** | 70.89 | 22.87 | 34.56 | 24.82 |
| **Naïve Bayes** | 92 | 24.7 | 39 | 96.29 |

**Table 10. Performance Measure for R2L Layer with Gain Ratio**

| | Precision (%) | Recall (%) | F-Value (%) | Accuracy (%) |
|---|---|---|---|---|
| **C5.0** | 100 | 100 | 100 | 100 |
| **MLP** | 66.67 | 2.11 | 4.1 | 88.21 |
| **Naïve Bayes** | 100 | 100 | 100 | 100 |

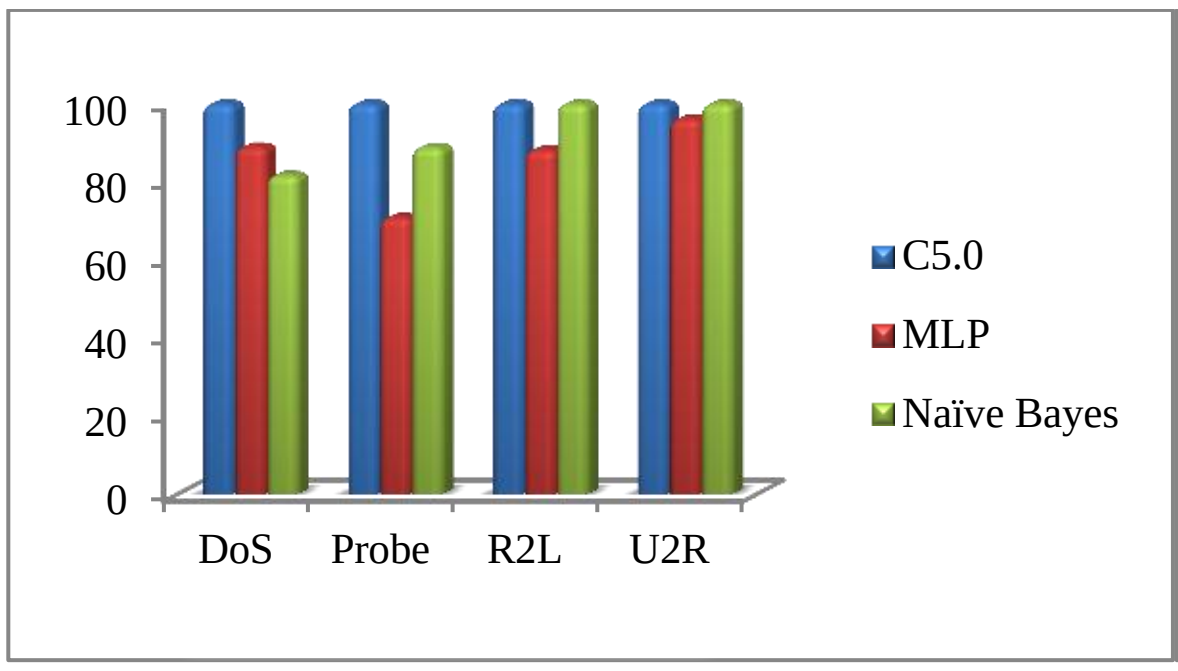

**Fig 2. Classification Rate (Accuracy) using Gain Ratio**

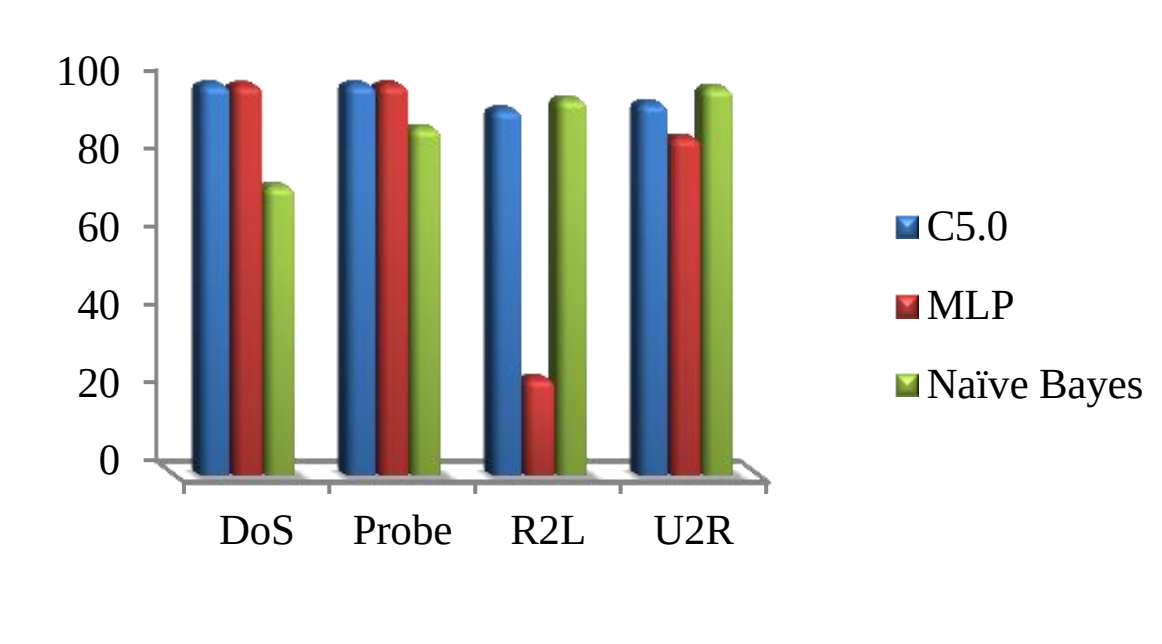

**Fig 3. Classification Rate (Accuracy) using 41 Features**

# 6. CONCLUSION AND FUTURE WORK

A multi-Layer intrusion detection system has been developed to achieve high efficiency and improve detection and classification rate accuracy. The proposed system consists of two stages. First stage is for attack detection and the second stage is for attack classification. The data is input in the first Stage which identifies if this record is a normal record or attack.

If the input record was identified as an attack then the administrator would be alarmed that the coming record is suspicious and then this suspicious record would be introduced to the second stage which consists of four sequential layers that specifies the class of this attack (DOS, probe, U2R or R2L). Finally the administrator would be alarmed of the expected attack type.

We examined each layer using different machine learning models (C5, MLP & Naïve Bayes) then we implemented our system with gain ratio feature selection technique for selecting the best features for each layer based on the attacks' type that the layer is trained to detect rather than using all the 41 features.

The advantage of the proposed mutli-layer system is not only the higher accuracy but also the multi-layers improve scalability as when new attacks of specific class are added to the dataset, there is no need to train all the layers but only the layer affected by the new attack. Attacks that are misclassified by the IDS as normal instances or given wrong attack class/type will be relabeled by the network administrator as the training module can be retrained at any point of time which makes its implementation adaptive to any new environment or any new attacks in the network .In addition, Our proposed system propagates errors as to simulate the real system and results be more accurate and real.

Our experimental results show that C5 is very effective in improving the attack detection rate and classification rate with low False Alarm Rate. Feature selection using Gain Ratio and implementing the Layered Approach reduce the time required to train and test the model significantly.

Most of the present methods for intrusion detection fail to reliably detect R2L and U2R attacks, while our proposed system can efficiently detect and classify such attacks.

The experimental results also show that C5 decision tree has significant detection and classification rate for both stages. Using Gain Ratio significantly enhances the accuracy of U2R and R2L for the three machine learning techniques (C5, MLP and Naïve Bayes). It was shown that MLP has high classification rate when using the whole 41 features in Dos and Probe layers.

The Future work can be directed towards finding ways to eliminate False alarm rate for MLP and Naïve Baise. Also using other Machine learning techniques in our experiments for detecting more types of intrusions. The layers sequence can be altered to see how it will affect in the accuracy of each layer. Finally, we can apply the new attacks and data partitioning techniques on our layered approach as in [27].